\documentclass[conference]{IEEEtran}
\IEEEoverridecommandlockouts
\usepackage{cite}
\usepackage{amsmath,amssymb,amsfonts}
\usepackage{algorithmic}
\usepackage{graphicx}
\usepackage{textcomp}
\usepackage{xcolor}
\usepackage{multirow}
\usepackage{float}
\usepackage{threeparttable}
\usepackage{acronym}
\newacro{KV}{key-value}
\newacro{LLM}{large language model}
\newacro{CIM}{compute-in-memory}
\newacro{LoRA}{low-rank adaptation}
\newacro{PE}{processing element}
\newacro{IPCN}{Inter-PE computational network}
\newacro{SRPG}{SRAM reprogramming and power gating}
\newacro{CT}{compute tile}
\newacro{DMAC}{multiply-accumulate operations on dynamic data}
\newacro{SMAC}{multiply-accumulate operations on static weights}
\newacro{RRAM-ACIM}{resistive RAM analog compute-in-memory}
\newacro{SRAM-DCIM}{static RAM digital compute-in-memory}
\newacro{NMC}{network main controller}
\newacro{OS FeFET}{oxide semiconductor ferroelectric field-effect transistor}
\newacro{NVM}{non-volatile memory}
\newacro{BEOL}{back‑end‑of‑line}
\newacro{FEOL}{front‑end‑of‑line}
\newacro{M3D}{monolithic 3D}
\newacro{ADC}{analog‑to‑digital converter}
\newacro{DRAM}{dynamic RAM}
\newacro{MVM}{matrix-vector multiplication}
\newacro{SAR}{successive approximation register}
\newacro{ISA}{instruction set architecture}
\newacro{KV}{key-value}
\usepackage{cleveref}
\crefname{figure}{fig.}{figs.}
\def\BibTeX{{\rm B\kern-.05em{\sc i\kern-.025em b}\kern-.08em
    T\kern-.1667em\lower.7ex\hbox{E}\kern-.125emX}}
\begin{document}

\title{CIMERA: \underline{C}ompute-in-\underline{I}nterconnect and \underline{Me}mory with \underline{R}econfigur\underline{a}ble Precision for LLM Inference\\
}

\author{\IEEEauthorblockN{Yue Jiet Chong\textsuperscript{\dag}\textsuperscript{a}, Yimin Wang\textsuperscript{\dag}\textsuperscript{b}, Wei Zhang\textsuperscript{\dag}\textsuperscript{b}, Xuanyao Fong\textsuperscript{a}}
\IEEEauthorblockA{Department of Electrical and Computer Engineering, National University of Singapore, Singapore \\
Email: \{jason.yj.chong, kelvin.xy.fong\}@nus.edu.sg\textsuperscript{a}, \{yimin.wang, wzhang\}@u.nus.edu\textsuperscript{b}}
\thanks{This work is funded in part by the National University of Singapore through the Microelectronics Seed Grant (FY2024); and in part by A*STAR under the RIE2030 Energy-aware Accelerated Computing program (Award H25-MSR3439).
}
\thanks{\textsuperscript{\dag} Authors contributed equally to this work.}
}

\maketitle

\begin{abstract}
\Acp{LLM} impose significant computational and memory demands, creating challenges for energy‑efficient inference across platforms ranging from data centers to power‑constrained edge devices. Weight precision plays a critical role in balancing inference accuracy, throughput, and energy consumption, while modern LLM workloads exhibit pronounced heterogeneity and tolerance that favors adaptive precision execution. This paper presents \textit{CIMERA}, a reconfigurable‑precision LLM inference accelerator that integrates compute‑in‑interconnect and memory to mitigate the memory wall and enable precision‑aware execution. Compared to Nvidia H100, \textit{CIMERA} delivers up to $25\times$ and $10\times$ higher energy efficiency for 1B and 13B models, respectively.
\end{abstract}


\section{Introduction}

\begin{figure}[t]
    \centering
    \includegraphics[width=1\linewidth]{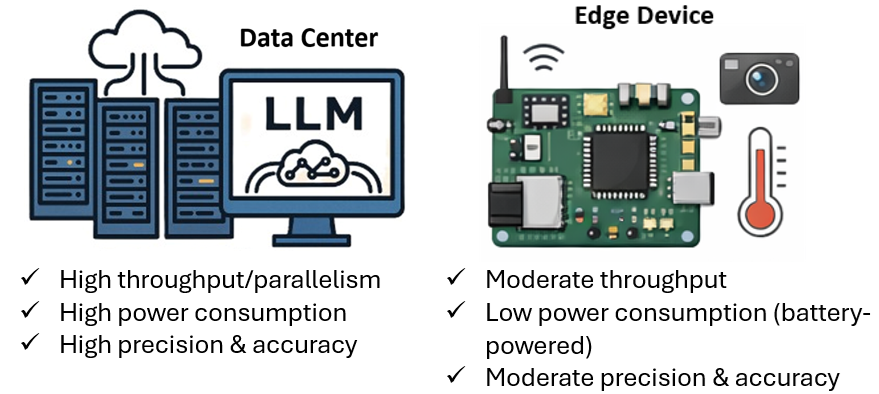}
    \caption{Distinct requirements of LLM deployment on data center and edge.}
    \label{fig:dc_vs_edge}
\end{figure}

As \ac{LLM}s are deployed across a wide spectrum of platforms from cloud-scale datacenters to power-constrained edge systems \cite{llm_in_industry}, there is a growing need for architectural techniques that can adapt model execution to diverse performance, accuracy, and energy requirements as shown in \Cref{fig:dc_vs_edge}.
Among the tunable hyperparameters, weight precision plays a critical role in determining the efficiency of \ac{LLM} inference.
While high-precision representations provide numerical robustness, they incur substantial storage and computation overhead.
Conversely, low-precision representations (e.g., INT4) significantly reduce memory footprint and energy consumption. 

Prior works \cite{llm_quant_1,llm_quant_2,llm_quant_3} have shown that LLMs exhibit strong tolerance to reduced precision, particularly when precision is selectively applied based on layer sensitivity, tensor type, or inference phase.
This observation has motivated the adoption of quantization and mixed-precision techniques as key enablers for efficient LLM deployment.
Modern LLM inference workflows exhibit significant phase-level and layer-level heterogeneity, such as the differing numerical requirements between prefill and decode stages, or between attention and feed-forward network (FFN) layers \cite{llm_hetero}.

Reconfigurable precision for model weights can be achieved through a combination of hardware and software co-design techniques.
On the algorithmic side, multi-bit quantization schemes, grouped quantization, and progressive precision scaling allow weights to be encoded and accessed at variable bit-widths.
On the hardware side, reconfigurable compute and memory structures enable dynamic selection of active bits during computation.
Compared to fixed-precision accelerators, such designs offer finer-grained control over energy–accuracy trade-offs and can better exploit workload-dependent redundancy in LLM computations.

In this work, we propose \textit{CIMERA}, an LLM inference accelerator architecture that integrates both in-memory and in-interconnect compute to mitigate the memory wall and support the deployment of LLM with weights of varying bit widths for aligning computational precision with workload requirements.

\section{CIMERA Hardware Architecture}
\begin{figure*}[t]
    \centering
    \includegraphics[width=1\linewidth]{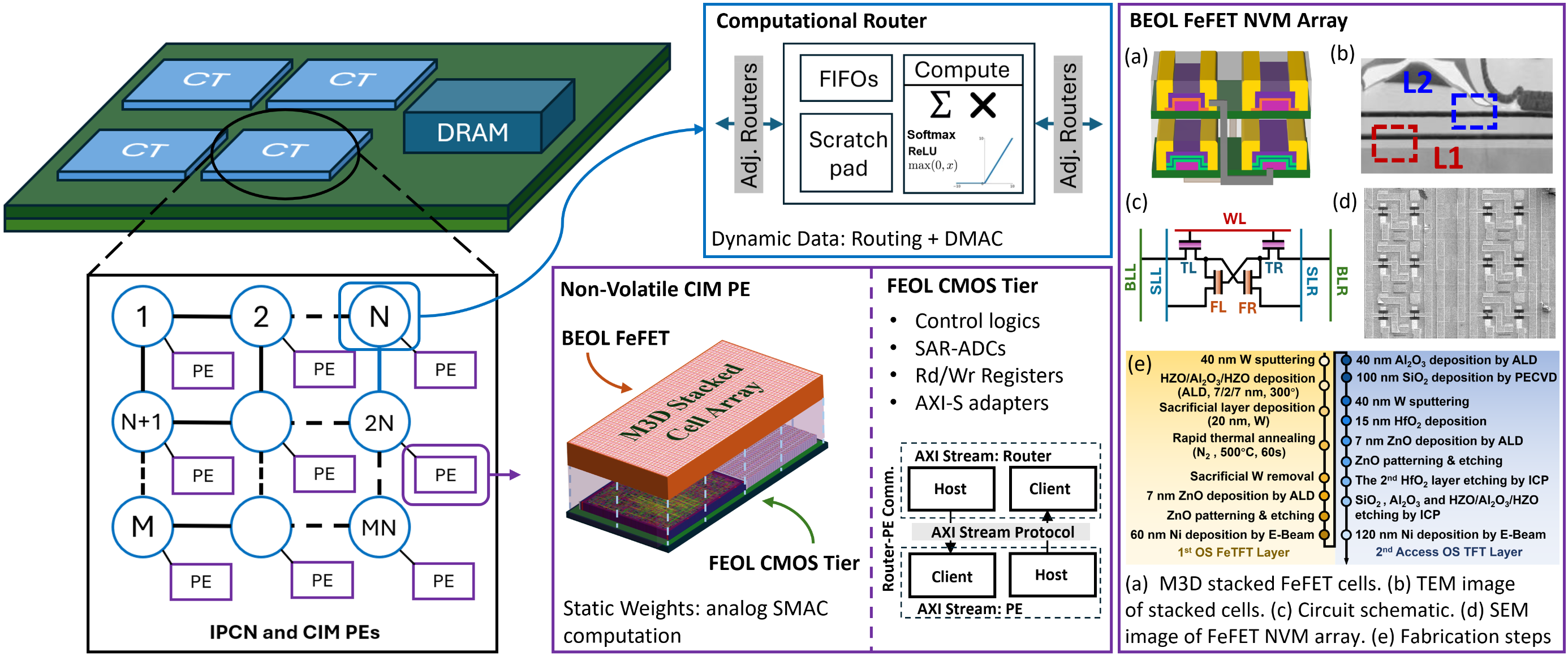}
    \caption{System Architecture of \textit{CIMERA}: IPCN, Router, CIM PE with FeFET NVM Array}
    \label{fig:cimera_sys_arch}
\end{figure*}

As \Cref{fig:cimera_sys_arch} shows, the \textit{CIMERA} hardware architecture adopts a chiplet-based design in which each chiplet contains \acp{PE} organized into a \ac{CT}.
Each \ac{CT} integrates a 2D-mesh \ac{IPCN} that interconnects multiple \ac{CIM} \acp{PE}.
The \ac{IPCN} orchestrates data movement across \acp{PE} and performs \ac{DMAC}, corresponding to the computation of attention scores ($\mathbf{Q} \cdot \mathbf{K^T}$).
In contrast, the \acp{PE} execute \ac{SMAC}, which implement the \acp{MVM} underlying the linear projections and feed-forward layers in the Transformer~\cite{transformer_arch} architecture.

\subsection{Processing Element (PE)}
In this work, \ac{OS FeFET} \ac{NVM} array (Fig.~\ref{fig:cimera_sys_arch}) is fabricated, characterized, and incorporated as \ac{PE} to perform \ac{SMAC}.
Each FeFET device stores a single neural network weight with up to 4‑bit precision (quad‑level cell, QLC), enabling compact multi‑bit weight representation within the memory array.
Spatial bit-slicing \cite{bit_slicing} across adjacent columns is adopted for bit width $>4$.
FeFETs intrinsically exhibit non‑volatile behavior by exploiting the remanent polarization of the ferroelectric gate dielectric, which modulates the channel conductance.
Due to the non‑volatility, model weights need to be programmed only once for a given \ac{LLM}, thereby eliminating weight reconfiguration and reducing runtime overhead.
Once initialized, the FeFET‑based \acp{PE} execute \ac{SMAC} directly in the analog domain, leveraging the massive parallelism of crossbar structures and the high energy efficiency of \ac{CIM}.

The FeFET offers strong compatibility with \ac{BEOL} fabrication processes \cite{ostft_beol_comptbl}.
FeFET can be fabricated after completion of the \ac{FEOL} CMOS logic transistors, enabling \ac{M3D} integration, in which FeFET memory layers are vertically stacked above conventional CMOS logic through sequential fabrication.
Consequently, all peripheral circuits, including control logic, \ac{ADC}, and registers are implemented in the base FEOL CMOS tier without incurring additional silicon area overhead.

\begin{figure}[t]
    \centering
    \includegraphics[width=1\linewidth]{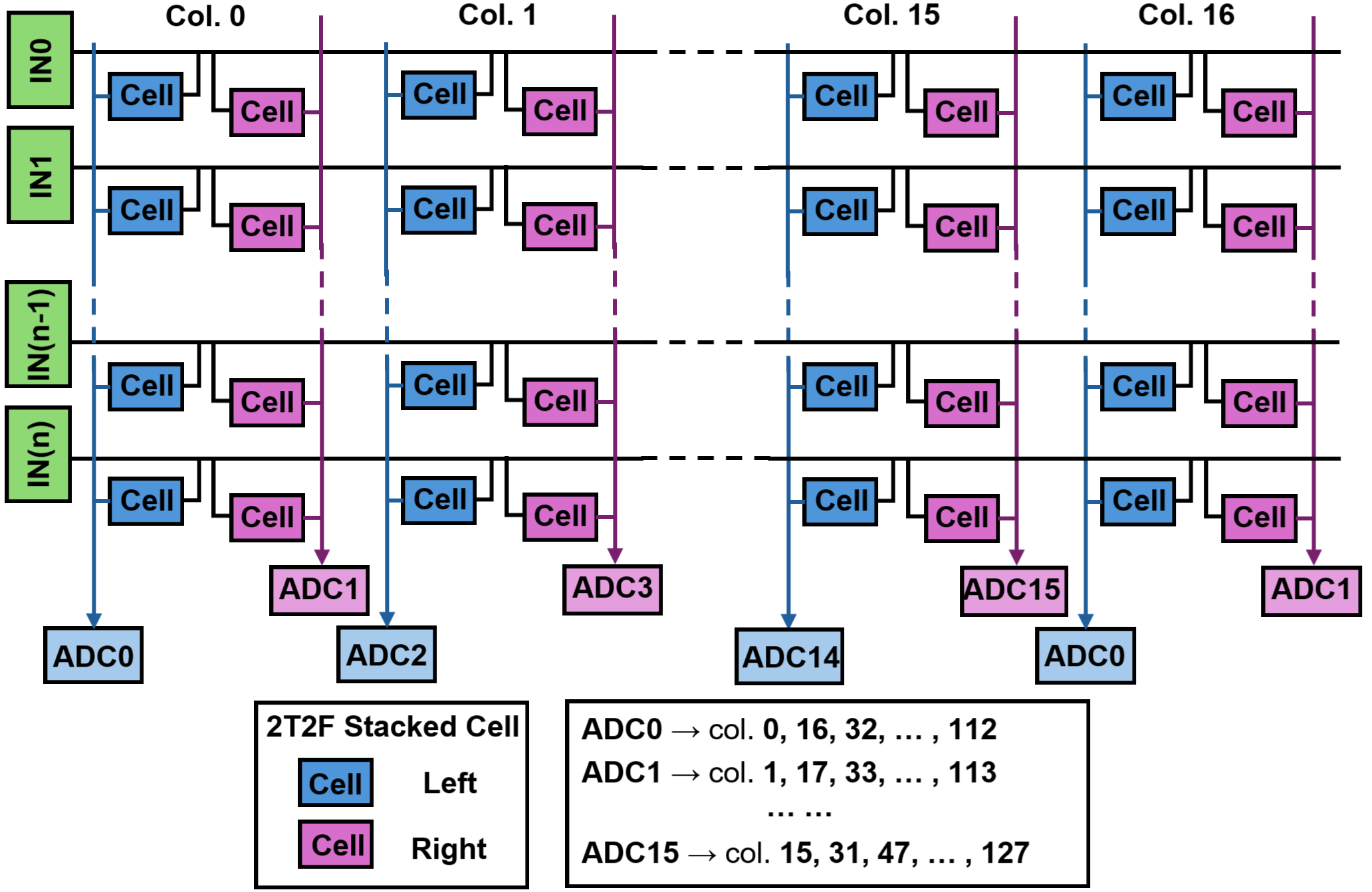}
    \caption{Schematic of FeFET \ac{NVM} Array with shared \acp{ADC}}
    \label{fig:nvm_array_sch}
\end{figure}

The \ac{SAR} \ac{ADC}~\cite{sar_adc} is adopted in the \ac{PE}.
It converts analog signals using an iterative binary search that sequentially determines each output bit via a capacitive DAC and a comparator.
This architecture enables reconfiguration of the conversion resolution by truncating the SAR loop, allowing the converter to operate with fewer decision cycles and reduced switching activity at lower precision, thereby achieving lower energy consumption and shorter conversion time.
The readout architecture employs 16 \acp{ADC} that are time-multiplexed across 128 memory columns using a strided sharing scheme (\Cref{fig:nvm_array_sch}).
In other words, $ADCi$ services columns $i+16k$ for $i \in \{0,\dots,15\}$ and $k=0,\dots,7$.
This organization enables uniform ADC utilization while reducing the overall ADC count.

\subsection{Inter-PE Computational Network (IPCN)}
The \ac{IPCN} is designed to seamlessly integrate \ac{CIM} \acp{PE} and to orchestrate efficient dataflow across the accelerator fabric for large‑scale \ac{LLM} inference.
In addition to serving as the on‑chip communication substrate, the \ac{IPCN} embeds computation to support key LLM workload primitives.
Operations such as \ac{DMAC}, partial‑sum reduction, and activation functions are executed directly within the network routers, thereby reducing data movement and improving overall execution efficiency.

To support diverse \ac{LLM} workloads and evolving execution patterns, the \ac{IPCN} incorporates a reprogrammable \ac{ISA} that provides fine‑grained control over data routing and in‑network computation.
Each router instance features four planar ports for inter‑router communication, along with two pairs of AXI‑Stream interfaces that connect to the \acp{PE}. Global coordination across the 2D‑mesh \ac{IPCN} is managed by the \ac{NMC}, which programs the routers using a predefined instruction stream stored in the instruction memory.

In transformer-based LLM inference, the \ac{KV} cache is used to store the key and value vectors generated by prior tokens, enabling reuse during the autoregressive decode phase and eliminating redundant recomputation of attention projections.
In \textit{CIMERA}, the SRAM scratchpad in each \ac{IPCN} router is utilized to store the corresponding \ac{KV} vectors for the layers mapped to the attached \ac{PE}.
By physically co-locating \ac{KV} cache storage with in-network computation and data routing, the architecture enables low-latency, high-bandwidth access to \ac{KV} vectors during attention score computation.

\section{Mapping and Dataflow Orchestration}
In practice, \ac{LLM} workloads exhibit structured yet irregular computation and communication patterns across layers, tensors, and inference phases.
Efficient execution on \textit{CIMERA} requires a mapping strategy that explicitly aligns model structure, data locality, and execution order with the architectural organization of the \acp{PE} and the \ac{IPCN}.
By jointly optimizing spatial mapping and dataflow at the software level, the system can minimize data movement, localize reductions, and maximize parallelism, as demonstrated in~\cite{iccad_leap, jetcas_jade}.

\subsection{Mapping}
The partitioned weight matrices $\mathbf{W_Q}$,$\mathbf{W_K}$,$\mathbf{W_V}$, and $\mathbf{W_O}$ are spatially distributed across the \ac{PE} crossbar arrays, while the corresponding intermediate tensors $\mathbf{Q}$, $\mathbf{K}$, $\mathbf{V}$, and $\mathbf{O}$ are allocated to distributed on‑chip scratchpad memory.
This separation allows weight matrices to remain stationary within the \acp{PE}, while intermediate data are accessed and reused efficiently during execution.
To further improve data locality, intermediate activations are co‑located with their corresponding weight matrices.
Specifically, each intermediate tensor is stored in the scratchpad memory associated with the \ac{IPCN} routers attached to the \acp{PE} hosting its weight matrix.
For example, the $\mathbf{KV}$ matrix is placed in the scratchpads of the routers neighboring the \acp{PE} where $\mathbf{W_K}$ and $\mathbf{W_V}$ are mapped.

\subsection{Dataflow}
The intra-layer dataflow of \ac{LLM} execution on \textit{CIMERA} is structured around three fundamental communication patterns, each corresponding to distinct computational phases of the Transformer model: \textit{broadcast}, \textit{reduction}, and \textit{unicast}.
These patterns are orchestrated by the \ac{IPCN} to efficiently match the data movement needs of attention computation and matrix–vector operations.
During the projection stage, input embeddings are broadcast to the \acp{PE} hosting the weight matrices, enabling parallel evaluation of the $\mathbf{QKV}$ projections across multiple \acp{PE}.
Since each weight matrix is spatially partitioned and distributed across multiple crossbar columns, the resulting partial \ac{SMAC} outputs are generated in parallel.
Unicast communication is employed to compute the attention scores using \ac{DMAC} within the \ac{IPCN}.

\begin{figure}[t]
    \centering
    \includegraphics[width=0.9\linewidth]{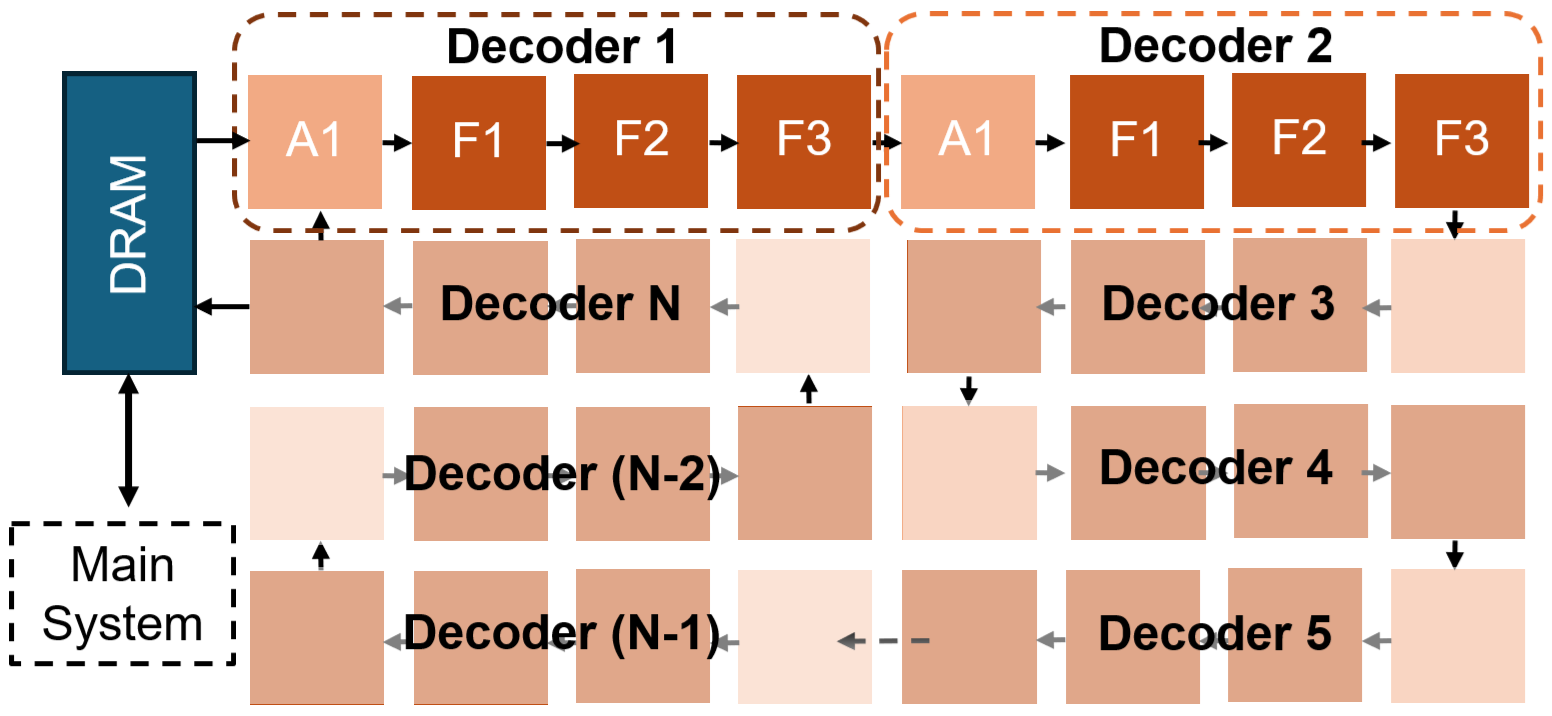}
    \caption{Serpentine-styled dataflow of \textit{CIMERA} (A: attention, F: feed-forward)}
    \label{fig:serpentine_dataflow}
\end{figure}

At the inter‑layer level, data propagation follows a serpentine execution topology as illustrated in \Cref{fig:serpentine_dataflow}, in which consecutive layers traverse across adjacent compute tiles.
This topology aligns with the layer‑by‑layer execution characteristic of transformer inference and enables efficient pipelining of computation and data movement across the accelerator fabrics.
The off‑chip DRAM serves as a centralized hub for external data exchange between the accelerator and the host system.
With the static placement of model weights within the \acp{PE}, off‑chip memory accesses are confined to the prefill stage, where input tokens are ingested, and to the end of the decode stage, where the generated output is returned to the host.

\begin{table}[t]
    \caption{System Parameters}
    \centering
    \setlength{\tabcolsep}{7pt}
    \begin{tabular}{c|c|c|c}
         \hline
         \multicolumn{4}{c}{\textbf{System Level}} \\
         \hline
         Bit-width & 64 & Frequency & 1 GHz \\
         \hline \hline
         \multicolumn{4}{c}{\textbf{Compute Tile Level}} \\
         \hline
         IPCN Dimension & 32$\times$32 & PE \# & 1024 \\
         \hline \hline
         \multicolumn{4}{c}{\textbf{Macro Level (per unit Router-PE pair)}} \\
         \hline
         PE Array Dimension & 128$\times$128 & ADC \# & 16 \\ 
         \hline
         Scratchpad Size & 32 KB & FIFO Size (each) & 128 B\\ 
         \hline
         DMAC \# & 16 & I/O Pairs \# & 6 \\ 
         \hline
    \end{tabular}    
    \label{tab:sys_param}
\end{table}

\section{System Evaluation}
The proposed system was evaluated using a comprehensive hardware–software co‑verification methodology, with parameters in Table~\ref{tab:sys_param}.
Digital hardware components were implemented and functionally verified using Verilog HDL.
Logic synthesis was performed with \textit{Synopsys Design Compiler}, followed by physical implementation using \textit{Cadence Innovus} for place‑and‑route.
The power and area of SRAM scratchpad macros were obtained using \textit{CACTI}~\cite{cacti}.
For the \ac{PE}, \ac{OS FeFET} memory cells and a prototype array were fabricated using \ac{BEOL}‑compatible \ac{M3D} integration \cite{ostft_beol_comptbl}.
Measurement results from the fabricated array were used to develop a compact model, which accurately captures the electrical characteristics and behavior of the array, and incorporated into the system‑level simulation framework.
End‑to‑end \ac{LLM} inference was evaluated using a cycle‑accurate, instruction‑level simulator that implements the \ac{IPCN} instruction set along with the proposed mapping and dataflow schemes \cite{iccad_leap}.
The simulator captures both computation and communication behavior across compute tiles, enabling detailed analysis of system‑level performance, energy efficiency, and scalability under representative \ac{LLM} workloads, as summarized in \Cref{tab:eval}.

\begin{table}[t]
    \caption{CIMERA Benchmarking: Throughput, Power, and Efficiency}
    \centering
    \setlength{\tabcolsep}{2.5pt}
    \begin{tabular}{c|c|c|c|c|c}
         \hline
         \multirow{2}{*}{Model}
         & Bit & Context Length & Throughput & Average & Efficiency \\
         & Precision & (Input/Output) & (tokens/s) & Power (W) & (tokens/J) \\ 
         
         \hline
         \multirow{6}{*}{\shortstack{LLaMA-3.2 \\ 1B}}
         & \multirow{2}{*}{4} & 1024/1024 & 1416.4 & \multirow{2}{*}{1.77} & 800.23 \\
         & & 2048/2048 & 789.4 & & 445.99 \\
         \cline{2-6}
         & \multirow{2}{*}{8} & 1024/1024 & 969.2 & \multirow{2}{*}{3.55} & 273.01 \\
         & & 2048/2048 & 566.4 & & 159.55 \\
         \cline{2-6}
         & \multirow{2}{*}{16} & 1024/1024 & 594.1 & \multirow{2}{*}{7.95} & 74.73 \\
         & & 2048/2048 & 361.9 & & 45.52 \\

        \hline
         \multirow{6}{*}{\shortstack{Mistral 7B}}
         & \multirow{2}{*}{4} & 1024/1024 & 520.5 & \multirow{2}{*}{5.31} & 98.02 \\
         & & 2048/2048 & 348.4 & & 65.61 \\
         \cline{2-6}
         & \multirow{2}{*}{8} & 1024/1024 & 309.8 & \multirow{2}{*}{10.95} & 28.29 \\
         & & 2048/2048 & 221.9 & & 20.26 \\
         \cline{2-6}
         & \multirow{2}{*}{16} & 1024/1024 & 171.2 & \multirow{2}{*}{23.01} & 7.44 \\
         & & 2048/2048 & 128.6 & & 5.59 \\

        \hline
         \multirow{6}{*}{\shortstack{LLaMA-2 \\ 13B}}
         & \multirow{2}{*}{4} & 1024/1024 & 333.9 & \multirow{2}{*}{7.97} & 41.89 \\
         & & 2048/2048 & 237.2 & & 29.76 \\
         \cline{2-6}
         & \multirow{2}{*}{8} & 1024/1024 & 192.4 & \multirow{2}{*}{16.81} & 11.45 \\
         & & 2048/2048 & 146.2 & & 8.70 \\
         \cline{2-6}
         & \multirow{2}{*}{16} & 1024/1024 & 104.1 & \multirow{2}{*}{33.63} & 3.10 \\
         & & 2048/2048 & 82.72 & & 2.46 \\

         \hline
    \end{tabular}
    \label{tab:eval}
\end{table}

\subsection{Performance Benchmarking}
\subsubsection{Throughput}
\begin{figure}[t]
    \centering
    \includegraphics[width=1\linewidth]{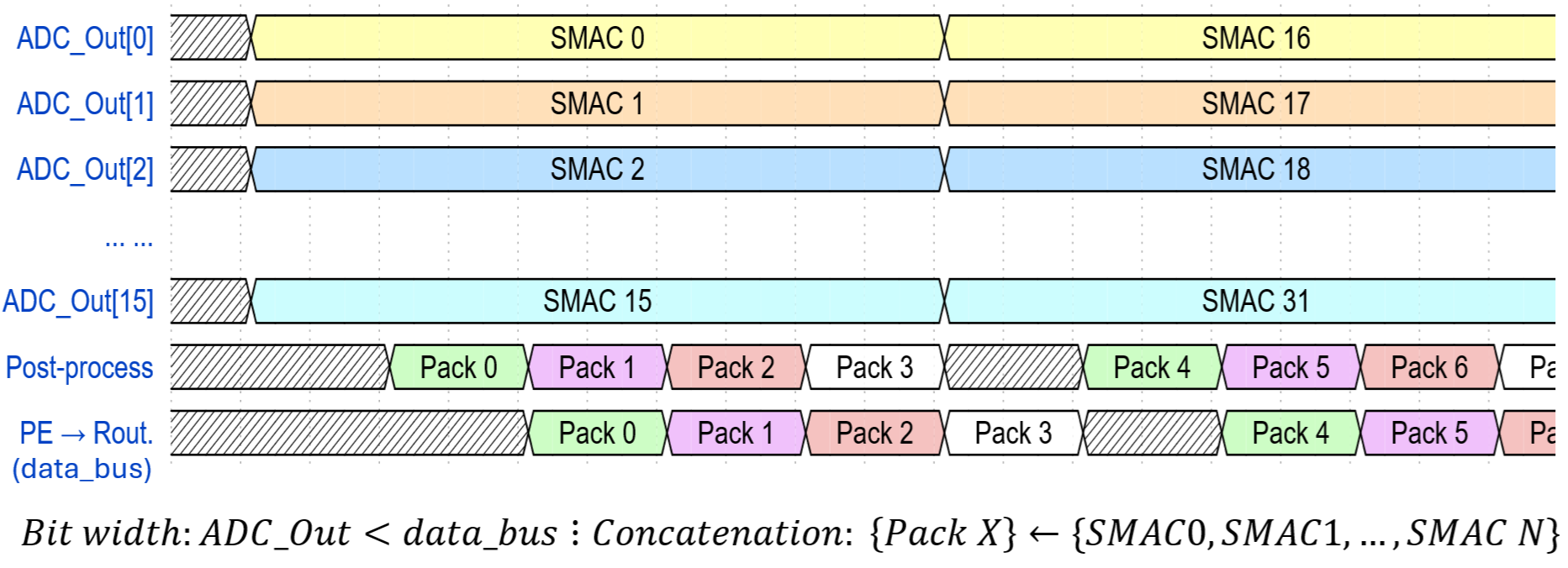}
    \caption{Timing diagram of PE operations}
    \label{fig:timing_diag}
\end{figure}

Across all evaluated models, lower weight precision consistently yields higher inference throughput. This throughput scaling is primarily attributed to the precision‑scalable \ac{SMAC} execution in the \acp{PE}.
At lower precision, fewer effective bit levels are activated during analog accumulation and peripheral conversion, thus reducing computation latency.
Moreover, reduced vector widths decrease communication pressure within the \ac{IPCN}, as more data can be concatenated and packed within each data packet (\Cref{fig:timing_diag}). 

As model size increases, the absolute throughput decreases due to a larger number of layers and increased inter‑layer traversal.
However, the relative throughput gain from precision reduction remains consistent, demonstrating the scalability of \textit{CIMERA}’s reconfigurable‑precision design.
Increasing the context length from 1024 to 2048 results in a throughput reduction across all precision settings.
This behavior is expected, as longer sequences increase attention computation and \ac{KV} cache access overhead.

\subsubsection{Power and Efficiency}
System power consumption increases with higher weight precision across all models and context lengths. Since model weights are stored statically in non‑volatile \acp{PE}, power scaling is dominated by computation and communication precision rather than repeated off-chip memory accesses.
Compared to fixed‑precision designs, \textit{CIMERA} allows sacrificing numerical fidelity for energy efficiency.
Moreover, with chiplet design and LLM workloads intrinsically executed in a sequential layer-by-layer manner, the power gating of idle chiplets enables sub-linear power scaling with respect to \ac{LLM} size.
Comparing to Nvidia H100, \textit{CIMERA} achieves better average energy efficiency across different \ac{LLM} sizes, from  1B (273~vs~11.2~tokens/J) to 13B (11.5~vs~1.2~tokens/J) at similar throughput.

\subsection{Macro Power and Area}
\Cref{tab:pwr_and_area} summarizes the power and area breakdown of the hardware macros.
The \ac{CIM} array in a \ac{PE} occupies the largest fraction of area as the \ac{OS FeFET} crossbar arrays integrate high‑density non‑volatile weight storage along with analog computation capability.
Although the \acp{ADC} are a major contributor to power, their area is not included in the planar footprint since they are implemented in the underlying \ac{FEOL} CMOS tier and the FeFET memory layers are vertically integrated above the \acp{ADC} using \ac{M3D} integration.
Thus, peripheral circuit complexity is effectively decoupled from array scaling.

\begin{table}[t]
    \caption{Avg. Power \& Area Breakdown of Hardware Macros (Unit)}
    \centering
    \setlength{\tabcolsep}{3.3pt}
    \begin{tabular}{c|cc|cc}
         \hline
         Macro$^\#$ & Power ($u$W) & Breakdown & Area (mm$^2$) & Breakdown \\
         \hline
         CIM Array & 120 & 3.45\% & 0.14 & 76.9\% \\
         \hline
         ADCs* & 3200 & 92.20\% & 0.074* & N/A \\
         \hline
         Scratchpad Mem. & 47 & 1.35\% & 0.013 & 7.2\% \\
         \hline
         Router & 105 & 3.00\% & 0.029 & 15.9\% \\
         \hline \hline
         Total & \multirow{2}{*}{3457} & \multirow{2}{*}{100\%} & \multirow{2}{*}{0.182} & \multirow{2}{*}{100\%} \\
         (Router-PE pair) & & & & \\
         \hline
    \end{tabular}

    \begin{tablenotes}
        \item $^\#$Technology node: 7~nm; Area per \ac{CT}: 186.4 mm$^2$
        \item *Total of 16 units, not contributing to area due to M3D
    \end{tablenotes}
    \label{tab:pwr_and_area}
\end{table}

\section{Conclusion}
By exploiting the inherent tolerance of \acp{LLM} to reduced and heterogeneous numerical precision, \textit{CIMERA} enables flexible trade‑offs among accuracy, throughput, and energy efficiency.
Through comprehensive hardware–software co‑verification and system‑level benchmarking, scalable improvements in both throughput and energy efficiency are demonstrated as weight precision is reduced, across models with varying context lengths and parameter counts.
Compared to fixed‑precision accelerators, \textit{CIMERA} provides a flexible architectural substrate that aligns computational precision with workload heterogeneity, enabling efficient deployment from edge devices to data‑center‑class systems.
It achieves better average energy efficiency across \ac{LLM} 1B (273~vs~11.2~tokens/J) to 13B (11.5~vs~1.2~tokens/J) at similar throughput as comparing to the Nvidia H100.
This demonstrates the effectiveness of scaling up \ac{CIM} with compute-in-interconnect.

\bibliographystyle{IEEEtran}
\bibliography{references}

\end{document}